\documentstyle[twocolumn,aps]{revtex}

\begin{document}

\twocolumn[\hsize\textwidth\columnwidth\hsize\csname
               @twocolumnfalse\endcsname


\title{Physical Interpretation of Cylindrically Symmetric 
Static Gravitational Fields}


\author{J. Colding$^\dagger$
 and
N. K. Nielsen$^\ast$\\
Fysisk Institute, Odense University, DK-5230 Odense M, Denmark\\
\vspace{0.2cm}
Y. Verbin$^\star $\\
Department of Natural Sciences, \\The Open University of Israel,
P.O.B. 39328, Tel Aviv 61392, Israel\\}

\bibliographystyle{unsrt}

\maketitle


\begin{abstract}
The explicit relationship is determined between the interior properties
of a static cylindrical matter distribution
and  the metric of the exterior space-time according to Einstein
gravity
for space-time dimensionality larger or equal to four.
This is achieved through use of a coordinate system isotropic in the
transverse coordinates. As a corollary,
similar results
are obtained for a spherical matter distribution in Brans-Dicke gravity
for dimensions larger than or equal to three.
The approach used here leads to consistency conditions for those parameters
characterizing the exterior metric. It is shown that these conditions  are
equivalent to the requirement
of hydrostatic equilibrium of the matter distribution
(generalized Oppenheimer-Volkoff equations). These conditions lead to a
consistent Newtonian limit where pressures and
the gravitational constant go to zero at the same rate. 
\end{abstract}

\vskip2pc]

\renewcommand{\theequation}{\arabic{section}.\arabic{equation}}

\section{Introduction}

Cylindrically symmetric space-times were first investigated in the
framework of Einstein's general relativity by Levi-Civita \cite{LC} and
Weyl \cite{Weyl} in 1917-1919.  Despite its age, the issue is still of
much interest and under  active research, especially since the
introduction of cosmic strings \cite{KibbleH} by  Kibble \cite{Kibble1},
Zel'dovich
\cite{Zel} and Vilenkin \cite{Vil1}.

The spacetime which is generated around a static cylindrically-symmetric
source is descried by a line element of the form
($\varrho$ is a radial coordinate; plain $r$ is reserved for future use):

\begin{equation}
ds^{2} = N_0({\varrho})dt^{2} - N_1({\varrho})d{\varrho}^{2} -
 N_2(\varrho)d{\phi}^2 - N_3(\varrho)dz^{2}.
\label{noname}
\end{equation}

The functional form of the metric components depends on the choice of
coordinate system (gauge). This freedom is the relic of the original
diffeomorphism symmetry. The two most popular gauges in the literature
\cite{exact-sol} are:\\
1) Weyl - Levi-Civita gauge: $N_1(\check{r})=N_3(\check{r})$ which gives
\begin{eqnarray}
ds^{2} =(k\check{r})^{2p}dt^{2} - (k\check{r})^{2p(p-1)}
(d\check{r}^{2}+dz^{2})
\nonumber \\
 - \gamma^{2} (k\check{r})^{-2p}\check{r}^{2}d{\phi}^2, \label{WLC}
\end{eqnarray}
where $p$ and $\gamma$ are free parameters and $k$ sets the length scale.
A variant of the Weyl - Levi-Civita
gauge is the Thorne gauge \cite{Thorne} which is defined by $N_1=N_0$.\\
2) Kasner gauge $N_1(\hat{r})=1$ which gives

\vspace{1 mm}

\begin{center}
\line(1,0){240}
\end{center}

\vspace{1 mm}

\noindent$ ^\dagger$Electronic address: colding@dirac.fys.ou.dk\\
$^\ast$Electronic address: nkn@fysik.ou.dk\\
$^\star$Electronic address: verbin@tavor.openu.ac.il

\newpage

\begin{eqnarray}
ds^{2} = (k\hat{r})^{2a}dt^{2} - (k\hat{r})^{2c}dz^{2} - d\hat{r}^{2}
\nonumber \\
 - \gamma^{2} (k\hat{r})^{2(b-1)}\hat{r}^{2}d{\phi}^2, \label{Kasner1}
\end{eqnarray}
where $a$, $b$, $c$ satisfy the Kasner conditions:
\begin{equation}
a + b + c=a^2 + b^2 + c^2=1. \label{Kasner2}
\end{equation}

$k$ and $\gamma$ are identical in both parametrizations only up to
a multiplicative constant. In what follows we will use only those that
appear in the Kasner metric.

It is obvious from the above forms of the line-element that the general
static cylindrically-symmetric vacuum solution of Einstein equations is
characterized by two free parameters. A question of fundamental interest
is then the interpretation of these two parameters and the connection
between them and the internal properties of the matter distribution.

A large amount of progress in this subject originated from the intensive
studies of cosmic strings, although the first results were obtained earlier
\cite{Marder1,Marder2,Bonnor}. It is well-known by now that the
parameter $\gamma$ describes a conic angular deficit which is also
related to the mass distribution of the source \cite{Marder2,Bonnor}.
Around a so called "gauge string" \cite{KibbleH} (i.e. one with ${\cal
T}^0_0={\cal T}^3_3$ as the
only non-vanishing components of the energy-momentum tensor), a simple
relation between the angular
deficit $\delta\phi = 2 \pi (1-\gamma)$ and the "inertial mass" (per unit
length) $\tilde{m}$, was found
\begin{equation}
\delta\phi = 8 \pi {\cal G}\tilde{m},
\label{angdef}
\end{equation}
first \cite{Vil1} in the linearized approximation assuming also an
infinitesimally thin source, and then \cite{Gott,Hiscock} by solving the
full non-linear Einstein equations around a uniform source (constant
${\cal T}^0_0$) with a finite radius. The same relation was also derived
\cite{Linet1} for a non-uniform source. Since the space-time around a
gauge string is locally flat ($p=0$ in (\ref{WLC}) or $a=c=0 , b=1$ in
(\ref{Kasner1})), this angular deficit is the only geometrical evidence
of its existence.

Further study of the subject \cite{Garfinkle1} involved a more realistic
model (i.e. the abelian Higgs model) for the gauge strings and the
analysis of the full coupled field equations for the gravitational field
and matter (scalar + vector) fields.

A major contribution to the understanding of the "relation between
gravitational mass, angular deficit and internal structure" was supplied
by Frolov, Israel and Unruh \cite{FIU} (FIU) who, without the use of any
specific model for the matter distribution, obtained several relations
between the parameters in the Kasner metric and integrals of the 
components of the energy-momentum tensor.

In the present work we take a model-independent approach.
We consider an arbitrary smeared matter distribution with cylindrical symmetry
and a finite radius $r_b$.
We start by introducing a line element which is isotropic
in the transverse coordinates \cite{mukh}, i.e. we take $N_2(r)=r^{2}N_1(r)$ in
(\ref{noname}). The line element is:
\begin{equation}
ds^{2} = e^{A}dt^{2} - e^{B}\delta_{ij}dx^i dx^j - e^{C}dz^{2}
\label{isotropic}
\end{equation}
where $i, j$ label the transverse coordinates, and $A$, $B$ and $C$ are
functions of the radial coordinate $r$ only.

This metric is similar to the one used by
Vilenkin \cite{Vil1}, although more general. It has the advantage
that the Einstein equations can be integrated by quadrature,
giving the functions $A$, $B$ and $C$  in terms of integrals of the components
of the energy-momentum tensor. These functions are subject to  a
consistency condition in the general cylindrically
symmetric case, which is
a manifestation of the hydrostatic equilibrium of the
source. The consistency condition is
studied in the Newtonian limit and it turns out that
the pressures and the gravitational constant go to zero at the same rate
with the masses kept
fixed. An exception is a gauge string, where the consistency condition
is trivially satisfied because the string tension is equal to the mass
density. Since it is both straightforward and instructive to generalize
the discussion to higher dimensions, we analyze the case of
$(D+1)$-Einstein gravity with $D>3$. We furthermore find analogous consistency
conditions in spherically symmetric space-time in $D$-dimensional
Brans-Dicke gravity, with $D\geq 3$. We finally use our results to show
that $D=3$ Brans-Dicke gravity has a Newtonian limit (unlike Einstein 
gravity). This last point may be regarded as a generalization of an 
earlier study \cite {Verbin}.

We also establish a connection between some of our results and those of FIU
\cite{FIU},
correcting on
the way some misprints which make their results difficult to use.

The outline of the paper is the following: In sec. II
we consider cylindrically symmetric solutions of
the Einstein equations for four-dimensional space-time ($D=3$). In sec. III
we discuss the interpretation of these solutions,
focussing on cosmic strings and the Newtonian limit. In sec. IV  we
generalise  to Einstein gravity in $D+1$ dimensions, $D>3$,
and in sec. V to Brans-Dicke gravity. Our results are summarized in sec. VI.

\setcounter{equation}{0}
\section{Cylindrically Symmetric Solutions in $D=3$}

With the line element (\ref{isotropic}),
the components of the Ricci tensor are in four space-time dimensions ($D=3$):
\begin{eqnarray}
{\cal R}_{00} &=& - \frac{1}{2} e^{\frac{A-C}{2}-B} \frac{1}{r} \frac{d}{dr} (r
A'e^{\frac{A+C}{2}}), \label{AApart}
\end{eqnarray}
\begin{eqnarray}
{\cal R}_{zz} &=& \frac{1}{2} e^{\frac{C-A}{2}-B} \frac{1}{r}
\frac{d}{dr} (rC'e^{\frac{A+C}{2}}) \label{CCpart}
\end{eqnarray}
\begin{eqnarray}
{\cal R}_{ij}
&=& \frac{1}{2} \delta_{ij}e^{- \frac{A+C}{2}} \frac{1}{r} \frac{d}{dr}[(rB'+2)
e^{\frac{A+C}{2}}]\nonumber \\
&+&\frac{x_i x_j}{2r^2}[2e^{- \frac{A+C}{2}}
\frac{1}{r} \frac{d}{dr}( r\frac{d}{dr}(e^{\frac{A+C}{2}}))\nonumber \\
&-& (B'+\frac{2}{r})(A'+C')-A'C'].  \label{Rij}
\end{eqnarray}
The source is described by the energy-momentum tensor with the
following components:
\begin{eqnarray}
        {\cal T}_{00} = \rho e^A \nonumber \\
        {\cal T}_{zz} = p_z e^C  \nonumber \\
        {\cal T}_{ij} = (p_r \frac{x^i x^j}{r^2} +
                    p_{\perp} \delta^{\perp}_{ij}) e^B
 \end{eqnarray}
with $\delta_{ij}^{\perp }=\delta_{ij}-\frac{x^i x^j}{r^2}$,
and $\rho $, $p_z$, $p_r$ and
$p_{\perp}$ are functions of the radial coordinate $r$ only.

It turns out to be convenient to use Einstein equations in the
form:
\begin{equation}
{\cal R}_{\mu\nu} = -8 \pi {\cal G}({\cal T}_{\mu \nu} -
\frac{1}{2} g_{\mu \nu}{\cal T})
\end{equation}
where ${\cal G}={\cal G}_4$ is the four-dimensional gravitational
constant and ${\cal T}$ is the contracted energy-momentum tensor:
${\cal T} = \rho - p_z - p_r - p_{\perp }$

By insertion of ${\cal T}_{\mu\nu}$ and use of
(\ref{AApart}) and (\ref{CCpart}) one obtains:
\begin{eqnarray}
e^{-B -\frac{A+C}{2}} \frac{1}{r} \frac{d}{dr}
(rA'e^{\frac{A+C}{2}})
\nonumber\\
=
8 \pi {\cal G}(\rho + p_z + p_r + p_{\perp })
\label{Einst0}
\end{eqnarray}
\begin{eqnarray}
e^{-B -\frac{A+C}{2}} \frac{1}{r} \frac{d}{dr}
(rC'e^{\frac{A+C}{2}})
\nonumber\\
= - 8 \pi {\cal G}(\rho + p_z - p_r - p_{\perp }).
\label{Einst3}
\end{eqnarray}
The ${\cal R}_{\phi\phi}$ equation is obtained from (\ref{Rij})
\begin{eqnarray}
e^{-B -\frac{A+C}{2}} \frac{1}{r} \frac{d}{dr}[(rB' + 2)
e^{\frac{A+C}{2}}]  \nonumber \\
= - 8 \pi {\cal G}(\rho + p_{\perp } - p_r - p_z),
\label{Einstphi}
\end{eqnarray}
and as a fourth equation we take the following combination:
\begin{eqnarray}
(B'+\frac{2}{r})(A'+C')+A'C'= 32 \pi {\cal G} e^B p_r .
\label{lastPart}
\end{eqnarray}

In vacuum the right-hand-sides of these equations vanish and the
first three of them are trivially integrated. In this way we may get a
line element equivalent to (\ref{WLC}) and (\ref{Kasner1}). The line element
contains
in the new version also two free parameters.
However, in order to gain some insight as to the meaning of the parameters
 and the relations between them, we need to
take into account, while solving the Einstein equations, the existence of
the source. One way to do it is to solve Einstein equations inside the
source ($r\leq r_b$) and join the interior solution with the exterior
vacuum solution on the boundary ($r=r_b$) using the formalism
of junction conditions \cite{Israel1}. This strategy requires some
assumptions concerning the matter in the source (i.e. its energy-momentum
tensor). However, by the special choice we made here for the coordinate
system, the Ricci tensor components have a form which is ready for
integration irrespective of the internal structure of the source,
as seen from (\ref{Einst0}), (\ref{Einst3}) and (\ref{Einstphi}). We
therefore exploit this fact and integrate the field equations in a way
that gives a physical meaning to the parameters in the exterior metric.

For that purpose we define:
\begin{eqnarray}
M(r)=2\pi\int _{0}^r r' dr' e^{\frac{A+2B+C}{2}}(\rho +p_r+p_{\perp } +p_z),
\label{M(r)}
\end{eqnarray}
\begin{eqnarray}
W(r)=-2\pi\int _{0}^r r' dr' e^{\frac{A+2B+C}{2}}(\rho -p_r+p_{\perp } -p_z).
\label{W(r)}
\end{eqnarray}
\begin{eqnarray}
X(r)=-2\pi\int _{0}^r r' dr' e^{\frac{A+2B+C}{2}}(\rho -p_r-p_{\perp } +p_z),
\label{X(r)}
\end{eqnarray}
We integrate (\ref{Einst0}), (\ref{Einst3}) and (\ref{Einstphi})
to obtain:
\begin{eqnarray}
rA'e^{\frac{A+C}{2}}=4{\cal G}M,
\label{AAAA}
\end{eqnarray}
\begin{eqnarray}
(rB' + 2)e^{\frac{A+C}{2}}-2=4{\cal G}W
\label{BBBB}
\end{eqnarray}
\begin{eqnarray}
rC'e^{\frac{A+C}{2}}=4{\cal G}X
\label{CCCC}
\end{eqnarray}
that by insertion into (\ref{lastPart}) lead to the equation
\begin{eqnarray}
(\frac{1}{2}+{\cal G}W)(M+X)+{\cal G}MX
=2\pi r^2p_re^{A+B+C}.
\label{hydrost}
\end{eqnarray}

The physical origin of (\ref{hydrost}) is the requirement of
hydrostatic equilibrium of the source which is necessary for having a
static solution.
It is essentially a first integral of the equation of
energy-momentum conservation as we can readily see by calculating
the derivatives
of both sides, using the other three field equations
 and getting rid of as
many metric components as possible, using
(\ref{AAAA}), (\ref{BBBB}) and (\ref{CCCC}).
We can then cast the conservation equation
into the form:
\begin{eqnarray}
e^{\frac{A+C}{2}}\frac{dp_r}{dr}&+&\frac{p_r-p_{\perp }}{r}
\nonumber \\
=
-\frac{2{\cal G}}{r}(M\rho
-Xp_z&+&(M+W+X)p_r-Wp_{\perp} ).
\label{Xcons}
\end{eqnarray}
This equation is the condition for hydrostatic equilibrium with
general-relativistic corrections. It may deserve the name
Oppenheimer-Volkoff equation (see e.g. MTW \cite{MTW} p. 605) for
cylindrical symmetry.

Outside the source (at $r\geq r_b$) we can integrate
(\ref{AAAA}), (\ref{BBBB}) and (\ref{CCCC}) analytically, since
$M$, $W$ and $X$ here are constant.
We find:
\begin{eqnarray}
A(r) = \frac{2M}{M+X}
\ln{(1+2{\cal G}(M+X)\ln{\frac{r}{r_0}}}),
\label{notyetNewton}
\end{eqnarray}
\begin{eqnarray}
B(r) = \frac{2W+\frac{1}{{\cal G}}}{M+X}
\ln{(1+2{\cal G}(M+X)\ln{\frac{r}{r_0}}}) -
2\ln{\frac{r}{r_0}}, \label{notyetr}
\end{eqnarray}
\begin{eqnarray}
C(r) = \frac{2X}{M+X}
\ln{(1+2{\cal G}(M+X)\ln{\frac{r}{r_0}}}), \label{notyetEinstein}
\end{eqnarray}
where $r_0$ is an arbitrary length scale.
Having determined the functions $A$, $B$ and $C$ outside the source we
have a third form of the line element of a cylindrically
symmetric space-time which again depends on
two parameters.
The angular deficit is not described
in this coordinate system by a parameter which multiplies the $d\phi$
term in the line element (like $\gamma$ in (\ref{Kasner1})). It is
rather hidden now in $W$ as we will see shortly.

The logarithms have the argument
$$
1+2{\cal G}(M+X)\ln{\frac{r}{r_0}}.
$$
Here we could get rid of the term 1 by a ${\cal G}$-dependent
redefinition of the scale $r_0$.
This redefinition would, however, destroy the possibility of obtaining a
Newtonian limit.

Note that $M$, $W$ and $X$ are not independent for $r>r_b$,
but obey according to (\ref{hydrost}) the consistency condition
\begin{eqnarray}
        (\frac{1}{2}+{\cal G}W)(M+X)+{\cal G}MX
        =0. \label{hydrostII}
\end{eqnarray}

The four equations (\ref{notyetNewton}), (\ref{notyetr})
(\ref{notyetEinstein}) and (\ref{hydrostII}), summarize the relation
between the exterior
metric and the matter distribution of the source, or stated somewhat
differently, describe how much information about the source can be
inferred
from the exterior geometry. Since the exterior metric contains two
independent parameters it is clear that a distant observer will have to
be satisfied with only two quantities in order to characterize the source.
One such quantity is the "Tolman mass" (per unit length) $M$.
As the second parameter we suggest the parameter $X$, which is the
corresponding quantity associated with the $z$-direction.
Notice that  the right-hand sides of (\ref{notyetNewton}) and
(\ref{notyetEinstein}) go
into each other under the interchange of $M$ and $X$. The left-hand sides
$A$  and $C$ are both solutions of Poisson-like equations and can be
interpreted as two potential functions.

The
connection to the Kasner metric (\ref{Kasner1}) is given by the
transformation between the radial variables $\hat{r}$ and $r$:
\begin{equation}
k\hat{r}=(1+2{\cal G}(M+X)\ln \frac{r}{r_0})^
{1+\frac{2{\cal G}W+1}{2{\cal G}(M+X)}}
\label{rtransformation}
\end{equation}
The constants $a, b, c, \gamma$ and $k$ in the Kasner metric
(\ref{Kasner1}) are expressed by the new parameters:
\begin{eqnarray}
a=\frac{M}{M+W+X+1/(2{\cal G})},
\nonumber \\
b=\frac{W+1/(2{\cal G})}{M+W+X+1/(2{\cal G})},
\nonumber \\
c=\frac{X}{M+W+X+1/(2{\cal G})}.
\label{connection}
\end{eqnarray}
\begin {equation}
\gamma = 1 + 2{\cal G}(M+W+X).     \label{gamma}
\end {equation}
The parameters $a$, $b$ and $c$ obey the Kasner conditions (\ref{Kasner2}).
One is obvious, while the other is a consequence of (\ref{hydrostII}).

This last issue, namely the connection between the Kasner parameters and
the matter distribution of the source, has been considered already by
FIU \cite{FIU}. Their results were obtained using a somewhat more geometric
approach,
based on an identity involving the extrinsic curvature and the Ricci tensor.
It is, however, difficult to use their results due to some typographical
errors in the relevant equations, as well as the fact that they are not
given in an explicit form which clarifies the full dependence of the
Kasner parameters on the quantities $M$, $W$ and $X$.
For the sake of completeness we give here the correct form of eqs.
(17)-(19) of FIU \underline{using their notation}.\\
$\bullet$ eq. (17). The right hand side should read $\frac{1}{4} k(c-b)$.\\
$\bullet$ eq. (18) should be:
\begin{eqnarray*}
\int _{\Delta z =1} (T_{\rho}^{\rho}+T_{\phi}^{\phi}-
T_{z}^{z} + T_{t}^{t} ) (-g)^{1/2} d^{3} x = \frac{1}{2} kb
\end{eqnarray*}
$\bullet$ eq. (19) should be:
\begin{eqnarray*}
\int _{\Delta z =1} (T_{\rho}^{\rho}+T_{\phi}^{\phi}) (-g)^{1/2} d^{3} x
 = {\frac{1}{4}} k(1-a)
\end{eqnarray*}
Note the difference in notation: FIU use dimensionless coordinates, the
parameters $a$, $b$ and $c$ differ by a cyclic permutation, and their $k$
is our $\gamma$. It should be stressed that FIU's $k$ has an implicit
dependence on the integrals of $T_{\mu}^{\nu}$. This dependence can be
unveiled by use of a third independent relation for the Kasner parameter
$a$ which is absent from FIU's paper:
\begin{eqnarray*}
\int _{\Delta z =1} (T_{\rho}^{\rho}-T_{\phi}^{\phi}+
T_{z}^{z} + T_{t}^{t} ) (-g)^{1/2} d^{3} x = \frac{1}{2} (ka-1)
\end{eqnarray*}
Adding this last equation with FIU's (16) and the correct form of (18)
yields an expression for FIU's $k$ which is identical with (\ref{gamma}).
Their other relations are contained in (\ref{connection}).

\setcounter{equation}{0}
\section{Interpretation, Cosmic Strings and Newtonian limit}

In order to get some feeling of the physics of the solutions we first
look for the familiar limits of the Minkowski space-time and the gauge
string. The Minkowski metric should be obtained in the absence of source,
namely: $M=W=X =0$. In this case we may either solve the
field equations again or take the appropriate limits in
(\ref{notyetNewton})-(\ref{notyetEinstein}).
Either way one finds immediately $A=B=C=0$.

The gauge string is a somewhat less trivial example. It is characterized by
$p_z = -\rho$ while $p_r=p_{\perp } =0$. Thus we
still have $A=C=0$. However, $W$ is now arbitrary (and yet
the condition (\ref{hydrostII}) is satisfied), so by taking the limit
$M \rightarrow 0$, $X\rightarrow 0$  in (\ref{notyetr}) we find that
$B=4{\cal G}W\ln{\frac{r}{r_0}}$. Therefore a gauge string is
described in our coordinate system by:
\begin {equation}
ds^{2} = dt^{2} - dz^{2} - (\frac{r_0}{r})^{-4{\cal G}W}(dr^{2} +
r^{2}d{\phi}^2)
\label{gaugestring1}
\end{equation}
This line element can be brought into a locally flat form by the
transformation $\frac{\hat{r}}{r_0} = \frac{1}{1+2{\cal G}W}
(\frac{r}{r_0})^{1+2{\cal G}W}$
which is a special case of (\ref{rtransformation}). This transformation
gives:
\begin {equation}
ds^{2} = dt^{2} - dz^{2} - d\hat{r}^{2} -
(1+2{\cal G}W)^{2} \hat{r}^{2} d{\phi}^2
\label{gaugestring2}
\end{equation}
which is the standard metric for a conical space-time with an angular
deficit of
\begin{equation}
\delta \phi =-4 \pi {\cal G}W.
\label{deltaphi}
\end{equation}
Using
(\ref{W(r)})
for the gauge string one gets Vilenkin's result (\ref{angdef}), with:
\begin{equation}
\tilde{m}
=2\pi\int _{0}^{r_b} r' dr' e^{\frac{A+2B+C}{2}}\rho .
\end{equation}
This is
also consistent with calculating
the Kasner parameters $a$, $b$, $c$ and
$\gamma$ by (\ref{connection}) and (\ref{gamma}).

Actually, it follows from this analysis that a conical
space-time
is generated in more general circumstances
than by a gauge string. The source may also have non-vanishing
$p_r$ and
$p_{\perp }$ provided their sum vanishes. In
this case we are dealing
with a cosmic string which is characterized by two parameters,
$\tilde{m}$ and say,
\begin{equation}
\tilde{p}_{r}
=2\pi\int _{0}^{r_b} r' dr' e^{\frac{A+2B+C}{2}}p_r.
\end{equation}
The angular deficit is still given by (\ref{deltaphi}), but
using here (\ref{W(r)}) for this special case we obtain a
generalization of Vilenkin's
result, namely:
\begin{equation}
\delta\phi = 8 \pi {\cal G}(\tilde{m} -\tilde{p}_{r}).
\label{deltaphi2}
\end{equation}

If we give up any restriction on $p_r$ and $p_{\perp }$,
but keep $p_z = -\rho$,  we have $X=M$, but not necessarily vanishing.
This makes (\ref{hydrostII}) reduce to
\begin{eqnarray}
        (1+2{\cal G}W)M+{\cal G}M^2
        =0. \label{hydrostIII}
\end{eqnarray}
One of the two solutions of (\ref{hydrostIII})  is
$
M=0
$
that by (\ref{AAAA}) and (\ref{CCCC}) again gives a conical
space outside the source, with
the deficit angle given by eq. (\ref{deltaphi2}).
If we choose the other possibility, viz. $X=M\neq 0$, $W=-\frac{1}{2}
(M+\frac{1}{{\cal G}})$,
the line element will still be
symmetric under boosts in the $z$-direction:
\begin {eqnarray}
ds^{2} = (1+ 4{\cal G}M \ln \frac{r}{r_0})(dt^{2} - dz^{2}) -\nonumber \\
-(1+4{\cal G}M \ln \frac{r}{r_0})^{-1/2} (\frac{r_0}{r})^{2}
(dr^{2} +r^{2} d{\phi}^2)
\label{closedspace}
\end{eqnarray}
This spacetime has the peculiar property that asymptotic azimuthal circles
have vanishing circumference. However, the matter distribution which
generates this solution is perfectly reasonable, so we may interpret this
solution as representing gravitationally collapsed cylindrical matter
distribution which is totally disconnected from the external space.
Actually, this solution has the same asymptotic behavior as the
Melvin universe\cite{Melvin}.
A special case of this situation is a  Higgs model cosmic string, which
was discussed  by
Garfinkle \cite{Garfinkle1}, who obtained
a  consistency condition equivalent to (\ref{hydrostIII}).

Another interesting case is $X=0$ which is possible for non-vanishing
$M$ if $W=-1/(2{\cal G})$. In that case we get
\begin {eqnarray}
ds^{2} = (1+ 2 {\cal G}M \ln \frac{r}{r_0})^2 dt^{2}-\nonumber \\
-(\frac{r_0}{r})^{2} (dr^{2} +r^{2} d{\phi}^2) - dz^{2}
\label{Kawaibh}
\end{eqnarray}
This is nothing but a trivial dimensional continuation of the Kawai solution
\cite{Kawai} which is actually a flat spacetime with cylindrical
topology\cite{Cornish1}. This solution may therefore be interpreted
not as an empty spacetime, but as the exterior solution around a special
kind of matter distribution confined to some cylindrical region in space.

Note that taking the limit $M \rightarrow 0$ in (\ref{closedspace}) or
(\ref{Kawaibh}) reproduces the extreme gauge string solution with
maximal angular deficit of $2\pi$.

Cosmic strings which generate conical space-time around them have the
unique property that they exert no force on non-relativistic matter.
This is obviously equivalent to the space-time being locally flat.
However, in the general case the solution exhibits curvature which in the weak
field
approximation manifests itself as a Newtonian potential
$V \simeq\frac{1}{2} A$.

In order to recover the Newtonian potential one would like to take the
limit of small
masses and pressures, with ${\cal G}$ fixed. However, the limiting procedure
must respect
(\ref{hydrostII}). Letting here e.g. $M$ and $X$ be small, we get
$W\simeq -1/(2{\cal G})$, which is inconsistent with the initial assumption.

A consistent way of defining a Newtonian limit is to keep masses fixed
and let
${\cal G}$  be small, with the pressures of order ${\cal G}$.
For small pressures
$W\rightarrow -M\hspace{0.1 mm}, X\rightarrow -M$, with $M$
having a fixed value (the Newtonian mass), while $M+X$,
which by (\ref{M(r)})
and (\ref{X(r)}) only involves the pressures in the
combination $p_r+p_{\perp }$,
has to be of order ${\cal G}$.
From
(\ref{hydrostII}) we get in this limit the consistency statement:
\begin{equation}
\frac{M+X}{2{\cal G}} \simeq-MX,
\label{Glimit}
\end{equation}
while
(\ref{notyetNewton}) implies:
\begin{equation}
V \simeq \frac{1}{2} A \simeq 2{\cal G}M\ln{\frac{r}{r_0}}
\label{Newton}
\end{equation}
which is recognized as the standard expression of the Newtonian potential
of a cylindrically symmetric mass distribution. It is curious that the
logarithmic form is
valid in a wider context than we are interested in: It is enough to take
$M+X \rightarrow 0$ in order to obtain
(\ref{Newton}).

To check the consistency of this limiting  procedure we consider
the
requirement of hydrostatic equilibrium (\ref{hydrost}),
or equivalently (\ref{Xcons}).
Keeping in (\ref{Xcons})
only terms of order ${\cal G}$, where pressures according to what was said
above
are already of order ${\cal G}$, we get
\begin{eqnarray}
\frac{dp_r}{dr}+\frac{p_r-p_{\perp }}{r}
\simeq-\frac{2{\cal G}M\rho}{r}
\end{eqnarray}
 which is the familiar condition for hydrostatic equilibrium
in Newtonian gravity.

\setcounter{equation}{0}
\section{Higher dimensionality}

In the previous sections an analysis of the static
solutions of the Einstein equations in 3+1 dimensions
for a general matter distribution with cylindrical symmetry
was carried out. The analysis
was facilitated by the use of a coordinate system where
the metric tensor is isotropic in the transverse coordinates.
It is natural to extend this analysis to higher dimensionalities,
in order to check which of the results we found still hold.

The metric tensor is still chosen isotropic in the transverse coordinates,
so the line element has the following form:
\begin{eqnarray}
      ds^{2} = e^{A}dt^{2} - e^{B}\delta_{ij}dx^i dx^j - e^{C}(dx^D)^{2},
        \nonumber \\
        \hspace{10 mm}i, j=1, \cdots D-1
\end{eqnarray}
with $A$, $B$ and $C$ functions of the radial coordinate $r$.
We define:
\begin{equation}
        \Upsilon = e^{\frac{A+(D-3)B+C}{2}}
        \label{Upsilon_def}
\end{equation}
The Ricci tensor has the following components:
\begin{eqnarray}
        {\cal R}_{00}
                = -\frac{1}{2} e^{A-B}
                \frac{1}{r^{D-2}}\Upsilon ^{-1}
                \frac{d}{dr} (r^{D-2} A'\Upsilon ),
                \label{d_R_00}
\end{eqnarray}

\begin{eqnarray}
        {\cal R}_{DD}
                = \frac{1}{2} e^{C-B}\frac{1}{r^{D-2}}\Upsilon ^{-1}
                \frac{d}{dr}( r^{D-2} C'\Upsilon).
                \label{d_R_33}
\end{eqnarray}

\begin{eqnarray}
        {\cal R}_{ij}
                =\delta_{ij}^{\perp }[\frac{1}{2}
                \frac{1}{r^{D-2}}\Upsilon ^{-1}
                \frac{d}{dr}( r^{D-2} B'\Upsilon)
                +\frac{1}{r} \Upsilon ^{-1}\Upsilon ']
                \nonumber \\
                +\frac{x^i x^j}{r^2}[\Upsilon ^{-1}\Upsilon ''
                +\frac{1}{2} \frac{1}{r^{D-2}}\Upsilon ^{-1}
                \frac{d}{dr}( r^{D-2} B'\Upsilon)
                \nonumber \\
                -\frac{1}{2} (D-2)(A'+ C')B'
                \nonumber \\
                 - \frac{1}{4} (D-2)(D-3)(B')^2 - \frac{1}{2} A'C'].
                \label{d_R_ij}
\end{eqnarray}

The energy-momentum tensor for a cylindrical distribution of energy and
pressure has the components:
\begin{eqnarray}
        {\cal T}_{00} = \rho e^A \nonumber \\
        {\cal T}_{DD} = p_D e^C  \nonumber \\
        {\cal T}_{ij} = (p_r \frac{x^i x^j}{r^2} +
                    p_{\perp} \delta^{\perp}_{ij}) e^B
 \end{eqnarray}
whence is found ${\cal T} = \rho -p_r- (D-2)p_{\perp}-p_D$.
 $\rho $, $p_D$, $p_r$ and
$p_{\perp}$ are functions of the radial coordinate $r$ only.

Einstein equations in $D+1$ dimensions imply:
\begin{equation}
    {\cal R}_{\mu \nu} = -\kappa_{D+1}({\cal T}_{\mu \nu} -
\frac{1}{D-1}g_{\mu \nu} {\cal T})
\label{equation}
\end{equation}
where
\begin{equation}
\kappa_{D+1} =\frac{D-1}{D-2}\Omega_{D-1}{\cal G}_{D+1}
\end{equation}
with $\Omega _{n}= 2\pi^{\frac{n+1}{2}}/\Gamma(\frac{n+1}{2})$ being
the area of the unit $n$-sphere and ${\cal G}_{D+1}$ Newton's constant
in $D+1$ dimensions.
The $00$ and $DD$ components of (\ref{equation}) are:
\begin{eqnarray}
        e^{-B} \Upsilon  ^{-1}
                \frac{1}{r^{D-2}}
                \frac{d}{dr} r^{D-2} A'\Upsilon
        \nonumber \\
        = \frac{2\kappa_{D+1}}{D-1}((D-2)\rho +p_r+(D-2)p_{\perp}+p_D)
\label{D_R_00},
\end{eqnarray}
\begin{eqnarray}
         e^{-B}\Upsilon ^{-1}
                \frac{1}{r^{ D-2}}
                \frac{d}{dr} r^{D-2} C'\Upsilon
         \nonumber \\
         = -\frac{2\kappa_{D+1}}{D-1}(\rho+(D-2)p_D-p_r-(D-2)p_{\perp})
\label{D_R_33},
\end{eqnarray}
The angular part is obtained from (\ref{d_R_ij}):
\begin{eqnarray}
        e^{-B}( \Upsilon ^{-1}
                \frac{1}{r^{ D-2}} \frac{d}{dr} r^{D-2} B'\Upsilon
        + \frac{2}{r}\Upsilon ^{-1}
        \frac{d}{dr}\Upsilon )
        \nonumber \\
         =- \frac{2\kappa_{D+1}}{D-1}(\rho-p_r+p_{\perp}-p_D),
        \label{mid_part}
\end{eqnarray}
By linear combination of (\ref{D_R_00}), (\ref{D_R_33})  and
(\ref{mid_part})
we obtain:
\begin{eqnarray}
        \frac{1}{r^{D-3}}\frac{d}{dr}r^{2D-5}\frac{d}{dr} \Upsilon
        = \kappa_{D+1}r^{D-2}e^{B} \Upsilon (p_r+p_{\perp })
        \label{second_Upsilon_def}
\end{eqnarray}
which in combination with the radial part of (\ref{equation})
is used to give as a fourth equation the following:
\begin{eqnarray}
        \frac{2(D-2)}{r}\frac{\Upsilon '}{\Upsilon }+\frac{1}{2}
        \frac{D-2}{D-3}[-\frac{1}{2}
         (A' + C')^2+2(\frac{\Upsilon'}{\Upsilon})^2]
         \nonumber \\
         +\frac{1}{2}A'C'=2\kappa _{D+1}e^Bp_r.
         \label{Cupsilon}
\end{eqnarray}

Now define the quantities:
\begin{eqnarray}
    U(r)=\frac{\kappa_{D+1}}{2(D-3)}\int_0^{r} (r')^{2D-5}e^{B}
    \Upsilon (p_r+p_{\perp })dr', \label{Y_0_N}
\end{eqnarray}
\begin{eqnarray}
    V(r)=-\frac{\kappa_{D+1}}{2(D-3)}\int_r^{r_b} r'e^{B}
    \Upsilon (p_r+p_{\perp })dr'. \label{V_0_N}
\end{eqnarray}
Eq.(\ref{second_Upsilon_def}) can be formally solved in terms of $U(r)$ and
$V(r)$:
\begin{equation}
        \Upsilon (r)= 1 -r^{2(3-D)}U(r)+V(r),
        \label{second_Upsilon_equation}
\end{equation}
where also the boundary condition
$\lim_{r \rightarrow \infty} \Upsilon = 1$ is used.

Defining furthermore the two quantities
\begin{eqnarray}
    M(r)&&=\Omega _{D-2}\int_0^{r} (r')^{D-2}e^{B}
    \Upsilon (\rho +\frac{p_D+p_r}{D-2}+p_{\perp})dr',
     \nonumber \\
    &&\label{a_0_N}
\end{eqnarray}
\begin{eqnarray}
    X(r)&&=-\Omega _{D-2}\int_0^{r} (r')^{D-2}e^{B}
    \Upsilon
    (\frac{\rho-p_r}{D-2}+p_D-p_{\perp})dr',
    \nonumber \\
    &&\label{b_0_N}
\end{eqnarray}
we can solve (\ref{D_R_00}) and (\ref{D_R_33}) since they imply
\begin{eqnarray}
        A' = \frac{\Omega _{D-1}}{\Omega _{D-2}}
        \frac{2{\cal G} _{D+1}M}{r^{D-2}\Upsilon }      \label{A'},
\end{eqnarray}
\begin{eqnarray}
        C' = \frac{\Omega _{D-1}}{\Omega _{D-2}}
        \frac{2{\cal G} _{D+1}X}{r^{D-2}\Upsilon }       \label{B'}.
\end{eqnarray}

Inserting (\ref{second_Upsilon_equation}), (\ref{a_0_N}) and (\ref{b_0_N})
into
(\ref{Cupsilon}) one gets:
\begin{eqnarray}
        2(D-2)(D-3)U(1+V)
         \nonumber \\
        +[\frac{\Omega _{D-1}}{\Omega _{D-2}}
        {\cal G} _{D+1}]^2
        (-\frac{1}{2}\frac{D-2}{D-3}(M+X)^2
        +MX)
        \nonumber \\
        =\kappa _{D+1}r^{2D-4}e^B\Upsilon ^2p_r. \label{QXZ}
\end{eqnarray}
As in the $D=3$ case, this equation is equivalent to a
generalized Oppenheimer-Volkoff equation. After some manipulations
on the same
line as in sec. II one gets the following differential equation
(cf.(\ref{Xcons})):
\begin{eqnarray}
        \Upsilon\frac{dp_r}{dr}
        +\frac{D-2}{r}(1+r^{2(3-D)}U+V)(p_r-p_{\perp})
        \nonumber \\
        =-\frac{\Omega _{D-1}}{\Omega _{D-2}}{\cal G} _{D+1}r^{2-D}
        (M\rho -Xp_D
        \nonumber \\
        -\frac{M+X}{D-3}(p_r-
        (D-2)p_{\perp}))
        \label{simple_consistency_equation}
\end{eqnarray}
The right-hand side of this equation is obtained from the right-hand side of
(\ref{Xcons})
using the substitution
$W\rightarrow -\frac{D-2}{D-3}(M+X)$.

An explicit expression for the metric tensor can again
be obtained for $r \geq r_b$
(outside the source), where $M$, $X$ and $U$ are constant, and we introduce
$r_h =U^{1/2(D-3)}$. These three quantities obey
according to (\ref{simple_consistency_equation}):
\begin{eqnarray}
        \left(\frac{\Omega _{D-1}}{\Omega _{D-2}}{\cal G}
_{D+1}\right)^2(-\frac{1}{2}\frac{D-2}{D-3}(M+X)^2+MX)
         \nonumber \\
        +2(D-2)(D-3)r_h^{2(D-3)}=0. \label{QYZ}
\end{eqnarray}
This consistency condition is the higher-dimensional generalization of
 (\ref{hydrostII}).
From (\ref{second_Upsilon_equation}), (\ref{A'}) and (\ref{B'}) we get
\begin{equation}
\Upsilon (r)=1-(\frac{r_h}{r})^{2(D-3)},
\end{equation}
\begin{equation}
A(r)=\frac{\Omega _{D-1}{\cal G}_{D+1}M}{(D-3)\Omega _{D-2}r_h^{D-3}}
\ln \frac{1-(\frac{r_h}{r})^{D-3}}{1+(\frac{r_h}{r})^{D-3}},
\label{ABS}
\end{equation}
\begin{equation}
C(r)=\frac{\Omega _{D-1}{\cal G}_{D+1}X}{(D-3)\Omega _{D-2}r_h^{D-3}}
\ln \frac{1-(\frac{r_h}{r})^{D-3}}{1+(\frac{r_h}{r})^{D-3}}
\label{CBS}
\end{equation}
that correspond to (\ref{notyetNewton})
and (\ref{notyetEinstein}) in the $D=3$ case.
The solution describes a black string with the horizon
at $r_h$. The line element
is again characterized by the two parameters $M$ and $X$.
The horizon parameter $r_h$ is determined from the properties of
the matter distribution through (\ref{Y_0_N}) and must obey the consistency
condition (\ref{QYZ}).
Inspection of the latter
equation shows that $r_h$ is always real.

From this solution, the spherical solution of Einstein gravity in
$D$ dimensions
can be obtained by taking $X=0$, in which case
the line element becomes, after a trivial dimensional reduction
(cf. (\ref{Kawaibh}))
\begin{eqnarray}
ds^2=\left(\frac{1-(\frac{r_h}{r})^{D-3}}
{1+(\frac{r_h}{r})^{D-3}}\right)^2dt^2
\nonumber \\
-\left(1+(\frac{r_h}{r})^{D-3}\right)^{\frac{4}{D-3}}
\delta_{ij}dx^i dx^j
\end{eqnarray}
that is the line element of a black hole in $D$ dimensions.
Schwarzschild coordinates are
obtained by the coordinate transformation
\begin{equation}
\hat{r}=\frac{1}{r}(r^{D-3}+r_h^{D-3})^{\frac{2}{D-3}}.
\end{equation}
In terms of $\hat{r}$ the line element is the Tangherlini solution
\cite{Tangherlini}
\begin{equation}
ds^2=\left(1-(\frac{r_h}{\hat{r}})^{D-3}\right)dt^2
-\frac{d\hat{r}^2}{1-(\frac{r_h}{\hat{r}})^{D-3}}-
\hat{r}^2d\Omega_{D-1}^2.
\end{equation}

In sec. III a thorough discussion of conical space-times
for $D=3$ was carried out. A parallel investigation for
$D>3$ is trivial: From (\ref{ABS}) and (\ref{CBS}) follows that $A=C=0$
requires $M=X=0$, and from the consistency condition (\ref{QYZ}) then
follows $r_h=0$, i.e. there is no angular deficit.

A consistent  Newtonian limit must respect the
consistency condition (\ref{QYZ}). Also in this case
this is obtained by letting pressures and the gravitational
constant go to zero at the same rate, with the mass kept fixed.
In this limit (\ref{simple_consistency_equation}) reduces to:
\begin{equation}
       \frac{dp_r}{dr}+\frac{D-2}{r}(p_r-p_{\perp })
       \simeq-\frac{{\cal G}_{D+1}M\rho }{r^{D-2}}
\end{equation}
which is recognized as the condition of hydrostatic equilibrium.
The Newtonian potential is then obtained from  (\ref{ABS}):
\begin{equation}
V\simeq \frac{1}{2} A=-\frac{\Omega _{D-1}{\cal G} _{D+1}M}{(D-3)\Omega _{D-2}}
        (\frac{1}{r})^{D-3}.
\end{equation}

\setcounter{equation}{0}
\section{Brans-Dicke gravity}

Brans-Dicke gravity in $D$ dimensions is obtained from Einstein gravity
in $D+1$ dimensions through the following steps:
\begin{enumerate}
\item the $D$-coordinate  is eliminated by
dimensional reduction,
\item
the combination
$$
\frac{1}{{\cal G}_{D+1}}e^{\frac{C}{2}},
$$
is identified with a scalar field,
\item a conformal transformation is carried out on the $D$-dimensional metric.
\end{enumerate}
This procedure leads to the following action:
\begin{equation}
        S = \int d^D x \sqrt{\mid g\mid } \left[\phi {\cal R}
          +\omega \frac{\nabla ^{\mu }\phi\nabla _{\mu }\phi}{\phi}
          +\frac{2(D-2)}{D-3}\Omega _{D-2} L\right].
        \label{BD_action}
\end{equation}
where the Lagrangian density of $D$-dimensional matter $L$ has been added to the
action. $\omega $ is a free parameter and $\phi $ a scalar field
\cite{BransDicke}.

Because of this connection between the two theories we expect an analogy
between the cylindrically
symmetric solutions of Einstein gravity in $D+1$ dimensions
discussed in the previous sections and
spherically symmetric solutions of Brans-Dicke gravity in $D$ dimensions.
Especially, results on
$D=3$ Brans-Dicke gravity can be obtained in this way.
To use this connection, we take again the line element in the isotropic form:
\begin{equation}
        ds^{2} = e^{A}dt^{2} - e^{B}\delta_{ij}dx^i dx^j
        \label{tutteli_du_2}
\end{equation}
with $A$ and $B$ functions of the radial coordinate $r$ only.

For $D=3$ the matter term is singular. For ordinary Einstein gravity, this
difficulty has been discussed by Cornish and Frankel \cite{CF}. They
suggest that it should be circumvented by use of a redefined gravitational
coupling constant ${\cal G}_R={\cal G}_3/(D-3)$ sacrificing the Newtonian
limit. In Brans-Dicke gravity this corresponds to using a redefined scalar
field $\Phi =(D-3)\phi$ with the difference that the Newtonian limit will
be kept. In the following we shall use the general formalism, ignoring
this complication, and return to it when specializing to $D=3$. 

The formal developments parallel those of secs. II and IV, with some
minor modifications. For this reason we omit many details,
giving only the important results.

The  energy-momentum tensor for a spherical matter distribution
has the same structure as those of the previous sections,
which means that it has components:
\begin{eqnarray}
        {\cal T}_{00} &=& \rho e^A \nonumber \\
        {\cal T}_{ij} &=& (p_r \frac{x^i x^j}{r^2} +
                    p_{\perp} \delta^{\perp}_{ij}) e^B
 \end{eqnarray}
whence ${\cal T} = \rho -p_r- (D-2)p_{\perp}$.
All quantities are functions of the radial coordinate $r$ only.

The resulting field equations are, as expected according to the
procedure sketched above for obtaining Brans-Dicke gravity, very
similar to those obtained for Einstein gravity in secs. II and IV,
with the substitution
$
e^{\frac{C}{2}}\rightarrow \phi .
$
The four independent equations corresponding to (\ref{D_R_00}), (\ref{D_R_33}),
(\ref{mid_part}) and (\ref{Cupsilon}) are:
\begin{eqnarray}
        &&\frac{1}{2}e^{-B}(\phi \Xi )  ^{-1}
                \frac{1}{r^{D-2}}
                \frac{d}{dr} (r^{D-2} A'\phi \Xi )
        \nonumber \\
        &&= \frac{D-2}{D-3}\frac{\Omega_{D-2}}{\phi}(\rho -{\cal T}
        \frac{\omega +1}{(D-2)\omega +D-1})
        \label{X_R_Y},
\end{eqnarray}
\begin{eqnarray}
        &&e^{-B}(\phi \Xi ) ^{-1}
                \frac{1}{r^{ D-2}}
                \frac{d}{dr} (r^{D-2} \phi '\Xi )
         \nonumber \\
         &&= -\frac{D-2}{D-3}\frac{\Omega _{D-2}}{(D-2)\omega +D-1}
        \frac{{\cal T}}{\phi }\label{X_R_X},
\end{eqnarray}
\begin{eqnarray}
        &&e^{-B}({\phi \Xi}) ^{-1}(\frac{1}{2}
                \frac{1}{r^{ D-2}} \frac{d}{dr}( r^{D-2}
               B'\phi \Xi )
        + \frac{1}{r}
        \frac{d}{dr}(\phi \Xi ))
        \nonumber \\
        && =- \frac{D-2}{D-3}\frac{\Omega_{D-2}}{\phi}(p_{\perp}
        +{\cal T} \frac{\omega +1}{(D-2)\omega +D-1}),
        \label{Xmid_partX}
\end{eqnarray}
and
\begin{eqnarray}
        e^{-B}(\frac{2(D-2)}{r} (\phi \Xi) ^{-1}(\phi \Xi)'
        &+&\frac{1}{2}((D-2)(A'+\frac{2\phi '}{\phi })B'
        \nonumber \\
        +\frac{1}{2}(D-2)(D-3)(B')^2&+&A'\frac{2\phi '}{\phi }
         -2\omega (\frac{\phi '}{\phi })^2)
        \nonumber \\
        = 2\frac{D-2}{D-3}\frac{\Omega_{D-2}}{\phi }
                p_r&&
        \label{second_Xi_def}
\end{eqnarray}
with
\begin{equation}
        \Xi = e^{\frac{A+(D-3)B}{2}}.
        \label{Xi_def}
\end{equation}

To proceed from here, one has to
distinguish between the cases $D>3$ and $D=3$.
First we consider $D>3$.
To obtain formal solutions of the differential equations we define the
quantities:
\begin{eqnarray}
    {\cal M}(r)=\Omega_{D-2}\int_0^{r} (r')^{D-2}e^{B}
    \Xi
    \times
   \nonumber \\
   (\rho-\frac{{\cal T} (\omega +1)}{(D-2)\omega +D-1})dr',
      \label{a_0_X}
\end{eqnarray}
\begin{eqnarray}
    {\cal X}(r)=-\frac{\Omega _{D-2}}{(D-2)\omega +D-1}&&
    \int_0^{r} (r')^{D-2}e^{B}\Xi {\cal T} dr',
     \nonumber \\
    && \label{b_0_X}
\end{eqnarray}
\begin{eqnarray}
    {\cal U}(r)=\frac{(D-2)\Omega _{D-2}}{2(D-3)^2}\int_0^{r} (r')^{2D-5}e^{B}
    \Xi (p_r+p_{\perp })dr', \label{YY_0_N}
\end{eqnarray}
\begin{eqnarray}
    {\cal V}(r)=-\frac{(D-2)\Omega _{D-2}}{2(D-3)^2}\int_r^{r_b} r'e^{B}
    \Xi (p_r+p_{\perp })dr'. \label{VV_0_N}
\end{eqnarray}
The solutions are then obtained from:
\begin{equation}
        \phi \Xi (r)= \phi _0 - r^{2(3-D)}{\cal U}(r)+{\cal V}(r)
        \label{second_Wpsilon_equation}
\end{equation}
with $\phi _0$ an integration constant, as well as
\begin{eqnarray}
        A' = \frac{2(D-2)}{D-3}
        \frac{{\cal M}}{r^{D-2}\phi \Xi }      \label{A''},
\end{eqnarray}
\begin{eqnarray}
        \phi ' = \frac{D-2}{D-3}
       \frac{{\cal X}}{r^{D-2}\Xi }       \label{B''}.
\end{eqnarray}
Corresponding to (\ref{QXZ}) one gets from (\ref{second_Xi_def}):
\begin{eqnarray}
        2(D-2)(D-3){\cal U}(\phi _0+{\cal V})
         \nonumber \\
        +(\frac{D-2}{D-3})^2
        (-\frac{1}{2}\frac{D-2}{D-3}({\cal M}+{\cal X})^2
        \nonumber \\
        +{\cal M}{\cal X}-\frac{\omega }{2}{\cal X}^2)
        =\frac{(D-2)\Omega _{D-2}}{D-3}r^{2D-4}e^B\phi \Xi ^2p_r
        \label{PXZ}
\end{eqnarray}
from which by differentiation a
generalized Oppenheimer-Volkoff equation is obtained:
\begin{eqnarray}
        \phi \Xi \frac{dp_r}{dr}
        +\frac{D-2}{r}(\phi_0+r^{2(3-D)}{\cal U}+{\cal V})(p_r-p_{\perp})
        \nonumber \\
        =-\frac{D-2}{D-3}r^{2-D}({\cal M}\rho
         -\frac{{\cal M}+(D-2){\cal X}}{D-3}p_r
        \nonumber \\
        +\frac{D-2}{D-3}
        ({\cal M}+{\cal X})p _{\perp }).
        \label{yimple_consistency_equation}
\end{eqnarray}

For $r \geq r_b$ the solution is given by:
\begin{eqnarray}
        \phi \Xi = \phi _0(1 -(\frac{r_h}{r})^{2(D-3)}),
\end{eqnarray}
\begin{eqnarray}
        A =\frac{(D-2){\cal M}}{2\phi _0(D-3)^2r_h^{D-3}}
        \ln \frac{1-(\frac{r_h}{r})^{D-3}}{1+(\frac{r_h}{r})^{D-3}},
        \label{BRA'}
\end{eqnarray}
\begin{eqnarray}
        e^{2\phi }=\frac{(D-2){\cal X}}{2\phi _0(D-3)^2r_h^{D-3}}
        \ln \frac{1-(\frac{r_h}{r})^{D-3}}{1+(\frac{r_h}{r})^{D-3}}
\label{BRfi'}
\end{eqnarray}
where the horizon radius $r_h=[\frac{{\cal U}}{\phi _0}]^{\frac{1}{2(D-3)}}$
is always real for $\omega >-\frac{D-1}{D-2}$ as a consequence of (\ref{PXZ}).

The gravitational constant is here proportional to $1/\phi _0$, and
it is seen from  (\ref{PXZ}) that a consistent
Newtonian limit is obtained by taking the pressures to zero at the
same rate as  $1/\phi _0$, keeping the mass fixed. The
Newtonian potential is then determined from  (\ref{BRA'}):
\begin{equation}
V\simeq \frac{1}{2} A=-\frac{(D-2){\cal M}}{2\phi _0(D-3)^2}
        (\frac{1}{r})^{D-3}.
\end{equation}
In the Newtonian limit (\ref{yimple_consistency_equation}) reduces to
the condition of hydrostatic equilibrium.

At $D=3$ we introduce $\Phi =(D-3)\phi $ and
obtain in analogy to (\ref{lastPart}), (\ref{AAAA}), (\ref{BBBB}) and
(\ref{CCCC}) in sec. II the following four equations:
\begin{equation}
(B'+\frac{2}{r})(A'+2\frac{\Phi'}{\Phi})+A'\frac{2\Phi '}{\Phi}
-2\omega (\frac{\Phi '}{\Phi })^2=\frac{16\pi e^{B} p_r}{\Phi },
\label{BRCons}
\end{equation}
\begin{equation}
rA'\Phi e^{\frac{A}{2}}=2{\cal M},
\label{calM}
\end{equation}
\begin{eqnarray}
(rB'&+&2)\Phi e^{\frac{A}{2}}-2\Phi(0)=2{\cal W},
\label{calW}
\end{eqnarray}
\begin{equation}
r\Phi ' e^{\frac{A}{2}}={\cal X}
\label{calX},
\end{equation}
with
\begin{equation}
{\cal M}(r)=2\pi \int _0^rr'dr'e^{\frac{A+2B}{2}}
(\rho -{\cal T} \frac{\omega +1}{\omega +2}),
\end{equation}
\begin{equation}
{\cal W}(r)
=-2\pi \int _0^rr'dr'e^{\frac{A+2B}{2}}
(p_{\perp }+{\cal T} \frac{\omega +1}{\omega +2}),
\end{equation}
and
\begin{equation}
{\cal X}(r)=-2\pi \int _0^rr'dr'e^{\frac{A+2B}{2}}
\frac{{\cal T}}{\omega +2}.
\end{equation}
Inserting (\ref{calM}), (\ref{calW}) and (\ref{calX}) into (\ref{BRCons})
it becomes
\begin{eqnarray}
({\cal W}+\frac{1}{2}\Phi(0))({\cal M}+{\cal X})+{\cal M}{\cal X}
-\frac{1}{2}\omega {\cal X}^2=
\nonumber \\
\pi r^2e^{A+B}\Phi p_r
\label{Xydrost}
\end{eqnarray}
that is equivalent to the generalized Oppenheimer-Volkoff equation:
\begin{eqnarray}
e^{\frac{A}{2}}\Phi\frac{dp_r}{dr}+\Phi (0)\frac{p_r-p_{\perp}}{r}=
\nonumber \\
-\frac{2}{r}({\cal M}\rho
+({\cal M}+{\cal W})p_r
-{\cal W}p_{\perp})
\label{Xtons}
\end{eqnarray}
Comparing this result with (\ref{yimple_consistency_equation}), we see
that the substitution
${\cal W}\rightarrow -\frac{D-2}{D-3}({\cal M}+{\cal X})$ gives the correct
right-hand side of (\ref{yimple_consistency_equation}).
At $r\geq r_b$ (outside the matter distribution) eq. (\ref{Xydrost}) gives
rise to  a consistency condition.

The solutions for $A$, $B$ and $\Phi $ are here, as seen from
(\ref{calM}), (\ref{calW}) and (\ref{calX}):
\begin{eqnarray}
A(r)&=& \frac{2{\cal M}}{{\cal M}+{\cal X}}
\ln{(1+\frac{{\cal M}+{\cal X}}{\Phi _0}\ln{\frac{r}{r_0}}}),
\label{notyetBRewton}
\end{eqnarray}
\begin{eqnarray}
B(r)
=\frac{2{\cal W}+\Phi (0)}{{\cal M}+{\cal X}}
\ln{(1+\frac{{\cal M}+{\cal X}}{\Phi _0}\ln{\frac{r}{r_0}}})
&-&2\ln{\frac{r}{r_0}}, \label{notyetBRr}
\end{eqnarray}
\begin{eqnarray}
\ln \frac{\Phi (r)}{\Phi (r_0)} =\frac{X}{{\cal M}+{\cal X}}
\ln{(1+\frac{{\cal M}+{\cal X}}{\Phi _0}\ln{\frac{r}{r_0}}}),
\label{notyetBREinstein}
\end{eqnarray}
where $\Phi _0$ is a constant of integration that will play the role of
inverse gravitational constant (up to a constant of proportionality).

A few special cases are worth considering. If we take ${\cal M}={\cal
X}=0$, ${\cal W}$
can be arbitrary, and we obtain a locally flat conical space characterized
by an angular
deficit:
\begin{equation}
\delta \phi=-2\pi \frac{2{\cal W}+\Phi (0)-\Phi _0}{\Phi _0}.
\end{equation}
For ${\cal M}\neq 0,\hspace{0.1 mm} {\cal X}=0$ the solution is the
Kawai solution \cite{Kawai,Cornish1}.

A consistent Newtonian limit is obtained according to (\ref{Xydrost}) by
letting pressures
get small at the same rate as $1/\Phi (0)$, keeping the mass fixed.
Requiring a flat-space metric in this limit, we see from (\ref{notyetBRr})
that also $1/\Phi _0$ must get small in the limit such that:
\begin{equation}
\frac{\Phi (0)}{\Phi_0}\rightarrow 1.
\end{equation}
Using this observation, we get from (\ref {notyetBREinstein})
the Newtonian gravitational potential $V$:
\begin{equation}
V\simeq \frac{1}{2}A\simeq \frac{{\cal M}}{2\Phi _0}\ln{\frac{r}{r_0}}.
\end{equation}
Eq. (\ref{Xtons}) reduces in this limit to the equation of
hydrostatic equilibrium.

\setcounter{equation}{0}
\section{Conclusion}

In this work we have studied in detail the relation between the gravitational
field and the matter distribution of a cylindrically-symmetric source in
Einstein as well as a spherically-symmetric source in Brans-Dicke gravity.

We have taken a model-independent approach and obtained relations
between the two free
parameters in the general cylindrically-symmetric solution of Einstein gravity
and the integrals of
components of the energy-momentum tensor.
The two parameters that characterize the source may be taken as the Tolman
mass and
the corresponding quantity associated with the axial direction.

We further report on consistency conditions, relating the masses and
pressures in cylindrically symmetric space-time in Einstein gravity
and interpret them as a manifestation of the hydrostatic equilibrium of the
source.
This was achieved through use of a coordinate system isotropic in the
transverse coordinates. In this coordinate system the
line element is symmetric by interchange of time and the direction
of the axial coordinate.

We generalized the discussion from four dimensions ($D=3$ in our
notation) to higher dimensions, and analyzed the case of
$(D+1)$-dimensional Einstein gravity. We furthermore found analogous
consistency conditions in spherically symmetric space-time in $D$-dimensional 
Brans-Dicke gravity. The consistency
conditions were studied in the Newtonian limit and it turned
out that the pressures and the gravitational constant (appropriately
defined for each of the theories) go to zero at the same rate with the
masses kept fixed. A remarkable detail is  the fact that the role of inverse
gravitational constant is played by an integration constant.\\

\vspace{0.5cm}
{\bf Acknowledgement:}  We are grateful to W. Israel for a helpful and
encouraging correspondence.


\begin{thebibliography}{999}

\bibitem{LC} T. Levi-Civita, Rend. Acc. Lincei {\bf 26}, 307 (1917);
{\bf 27}, 3, 183, 220, 240, 283, 343 (1918); {\bf 28}, 3, 101 (1919).
\bibitem{Weyl} H. Weyl, Ann. Phys. (Leipzig) {\bf 54}, 117 (1917).
\bibitem{KibbleH} T.W.B Kibble and M. Hindmarsh,
Rep. Progr. Phys. {\bf 58}, 477 (1995).
\bibitem{Kibble1} T.W.B Kibble, J. Phys. {\bf A9}, 1387 (1976).
\bibitem{Zel} Ya. B. Zel'dovich, Mon. Not. R. Astron. Soc. {\bf 192}, 663
(1980).
\bibitem{Vil1} A. Vilenkin, Phys. Rev. {\bf D23}, 852 (1981).
\bibitem{exact-sol} D. Kramer, H. Stephani, E. Herlt and M. MacCallum,
Exact Solutions of Einstein's Field Equations (Cambridge UP, Cambridge,
England 1980).
\bibitem{Thorne} K.S. Thorne, Phys. Rev. {\bf 138}, B251 (1965).
\bibitem{Marder1} L. Marder, Proc. Roy. Soc. London A {\bf244}, 524 (1958).
\bibitem{Marder2} L. Marder, Proc. Roy. Soc. London A {\bf252}, 45 (1959).
\bibitem{Bonnor} W.B. Bonnor, J. Phys. A {\bf12}, 847 (1979).
\bibitem{Gott} J.R. Gott, Astrophys. J. {\bf288}, 422 (1985).
\bibitem{Hiscock} W.A. Hiscock, Phys. Rev. {\bf D31}, 3288 (1985).
\bibitem{Linet1} B. Linet, Gen. Relativ. Gravit. {\bf 17}, 1109 (1985).
\bibitem{Garfinkle1} D. Garfinkle, Phys. Rev.{\bf D32}, 1323 (1985).
\bibitem{FIU} V.P. Frolov, W. Israel and W.G. Unruh, Phys Rev.{\bf D39},
1084 (1989).
\bibitem{Vil2} A. Vilenkin, Phys. Rev. Lett. {\bf 46}, 1169 (1981).
\bibitem{mukh} B. Mukherjee, Bull. Cal. Math. Soc. {\bf 30}, 95 (1938).
\bibitem{Verbin} Y. Verbin, Phys. Rev. {\bf D50}, 7318 (1994).
\bibitem{Israel1} W. Israel, Nuovo Cimento {\bf 44B}, 1 (1966);
{\bf 48B}, 463 (E) (1966).
\bibitem{Melvin} M.A. Melvin, Phys. Lett. {\bf 8}, 65 (1964).
\bibitem{Kawai} T. Kawai, Prog. Theor. Phys. {\bf 94}, 1169 (1995).
\bibitem{Cornish1} N.J. Cornish, {\em The black hole that went away},
gr-qc/9609016.
\bibitem{MTW} C.W. Misner, K.S. Thorne and J.A. Wheeler, Gravitation
(Freeman, San Francisco 1973).
\bibitem{Tangherlini} F.R.Tangherlini, Nuovo Cimento {\bf 27}, 636 (1962).
\bibitem{BransDicke} C. Brans and R.H. Dicke, Phys Rev {\bf 124}, 925 (1961).
\bibitem{CF} N.J.Cornish and N.E.Frankel, Phys. Rev. {\bf D43}, 2555 (1991).
\end{thebibliography}
\end{document}